\documentclass[preprintnumbers,amsmath,amssymb]{revtex4}

\usepackage{graphicx}
\usepackage{color}

\begin{document}

\title{Transverse electric plasmons in bilayer graphene}

\author{Marinko Jablan and Hrvoje Buljan}
\affiliation{Department of Physics, University of Zagreb, Bijeni\v{c}ka c. 32, 10000 Zagreb, Croatia}

\author{Marin Solja\v{c}i\'{c}}
\affiliation{Department of Physics, Massachusetts Institute of Technology, 77 Massachusetts Avenue, Cambridge MA 02139, USA}

\begin{abstract}
We predict the existence of transverse electric (TE) plasmons in bilayer graphene. 
We find that their plasmonic properties are much more pronounced in 
bilayer than in monolayer graphene, in a sense that 
they can get more localized at frequencies just below $\hbar\omega=0.4$~eV
for adequate doping values. This is a consequence of the perfectly nested 
bands in bilayer graphene which are separated by $\sim 0.4$~eV. 
\end{abstract}

\maketitle

Plasmons are self-sustained collective electron excitations 
which are of great interest both for fundamental physics and potential 
technological applications. Plasmon is a paradigmatic quantum many-body 
phenomenon studied in condensed matter physics \cite{PinesBook}. 
Closely related excitations are surface plasmons which hold promise as 
a possible tool for controlling light at subwavelength scales 
\cite{Barnes2003,Atwater2005,Yablonovitch2005,Aristeidis2005}, giving rise to 
the field of nanophotonics, and they also play an important role in metamaterials 
\cite{Veselago1968,Shalaev2007,Pendry2000,Smith2004};
plasmons in essentially two-dimensional (2D) structures are similar in this 
respect to surface plasmons. 
These reasons are a great motivation for studying plasmonic excitations 
and their properties in novel materials. Two such materials are 
monolayer \cite{Novoselov2004,CastroNeto} and bilayer (see e.g., 
\cite{CastroNeto,Novoselov2006}) graphene. 
Graphene is a 2D sheet made of carbon atoms organized in a 
honeycomb lattice \cite{Novoselov2004,CastroNeto}, whereas bilayer graphene 
consists of two such layers stacked on top of each other in a certain way 
\cite{CastroNeto,Novoselov2006}. While there are only a few 
studies of plasmons in bilayer graphene \cite{Wang2007,Borghi2009,Wang2010,Sensarma2010}, 
these collective excitations have attracted substantially more attention in 
monolayer graphene \cite{Wunsch2006,Hwang2007,Mikhailov2007,
Rana2008,Kramberger2008,Liu2010,Jablan2009}. 
Several years ago it was predicted that graphene, besides the ordinary longitudinal 
plasmons [transverse magnetic (TM) modes] \cite{Wunsch2006,Hwang2007,Mikhailov2007,Rana2008,Jablan2009}, 
also supports unusual transverse plasmons [transverse electric (TE) modes] 
\cite{Mikhailov2007}. These excitation are possible only if the imaginary part of 
the conductivity of a thin sheet of material is negative \cite{Mikhailov2007}. 
On the other hand, such a conductivity requires some complexity of the 
band structure of the material involved. For example, TE plasmons 
cannot occur if the 2D material possesses a single parabolic 
electron band. From this perspective, bilayer graphene, with its rich band 
structure and optical conductivity (e.g., see \cite{Nicol2008} and references therein), 
seems as a promising material for exploring the possibility of existence of TE plasmons. 
Here we predict the existence of TE plasmons in bilayer graphene.
We find that their plasmonic properties are much more pronounced in 
bilayer than in monolayer graphene, in a sense that the
wavelength of TE plasmons in bilayer can be smaller than in monolayer graphene 
at the same frequency.

Throughout this work we consider bilayer graphene as an infinitely 
thin sheet of material with conductivity $\sigma ({\bf q},\omega)$. 
We assume that air with $\epsilon_r=1$ is above and below bilayer graphene. 
Given the conductivity, by employing classical electrodynamics, one 
finds that self-sustained oscillations of the charge occur when 
(see \cite{Mikhailov2007} and references therein)
\begin{equation}
1+\frac{i\sigma({\bf q},\omega) \sqrt{q^2-\omega^2/c^2}}
{2\epsilon_0\omega}=0
\end{equation}
for TM modes, and 
\begin{equation}
1-
\frac{\mu_0 \omega i\sigma({\bf q},\omega)}
{2\sqrt{q^2-\omega^2/c^2}}=0
\label{TE}
\end{equation}
for TE modes. The TM plasmons can considerably depart from the light 
line, that is, their wavelength can be considerably smaller than that 
of light at the same frequency. For this reason, when calculating 
TM plasmons it is desirable to know the conductivity as a function of 
both frequency $\omega$ and wavevector ${\bf q}$. However, it turns out 
that the TE plasmons (both in monolayer \cite{Mikhailov2007} and 
bilayer graphene, as will be shown below) are quite close to the light 
line $q=\omega /c$, and therefore it is a good approximation 
to use $\sigma (\omega)=\sigma ({\bf q}=0,\omega)$. 
Moreover, these plasmons are expected to show strong polariton character, 
i.e., creation of hybrid plasmon-photon excitations. 
At this point it is worthy to note that if the relative permittivity of
dielectrics above and below graphene are sufficiently different, so that
light lines differ substantially, then TE plasmon will not exist 
(perhaps they could exist as leaky modes).

The conductivity $\sigma (\omega)=\Re \sigma (\omega)+i\Im \sigma (\omega)$
is complex, and plasmon dispersion is characterized by the imaginary part 
$\Im \sigma (\omega)$, whereas $\Re \sigma (\omega)$ determines 
plasmon losses, or more generally absorption of the sheet. 
From Eq. (\ref{TE}) it follows that the TE plasmons exist only if 
$\Im \sigma (\omega)<0$ \cite{Mikhailov2007}. 

\begin{figure}[htbp]
\centerline{
\mbox{\includegraphics[width=0.44\textwidth]{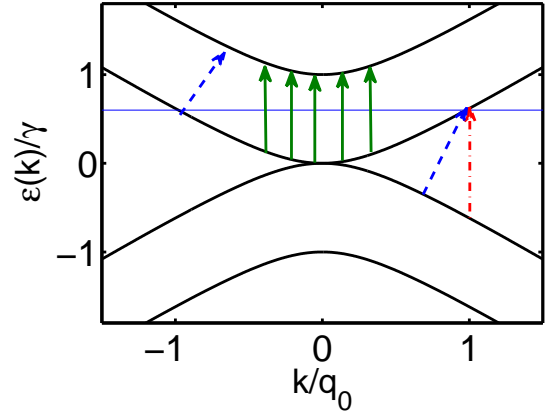}}
}
\caption{The band-structure of bilayer graphene. 
The two upper bands (as well as the two lower bands) are perfectly nested 
and separated by $\gamma\sim 0.4$~eV; $q_0=\gamma/\hbar v_F$. 
Horizontal line depicts one possible value of the Fermi level, and 
arrows denote some of the possible interband electronic transitions. 
See text for details. 
}
\label{bands}
\end{figure}

In order to calculate the imaginary part of the conductivity, we 
employ Kramers-Kronig relations and the calculation of absorption 
by Nicol and Carbotte \cite{Nicol2008}, where $\Re\sigma(\omega)$ 
[see Eqs. (19)-(21) in Ref. \cite{Nicol2008}]
was calculated by using the Kubo formula. 
The optical conductivity has rich structure due to the 
fact that the single-particle spectrum of graphene is organized 
in four bands given by \cite{Nicol2008}, 
\begin{equation}
\frac{\epsilon({\bf k})}{\gamma} =
\pm\sqrt{\frac{1}{4}+\left(\frac{\hbar v_F k}{\gamma}\right)^2}
\pm \frac{1}{2},
\label{bandstructure}
\end{equation}
where $v_F=10^6$~m/s, the parameter $\gamma\approx 0.4$~eV 
is equal to the separation between the two conduction bands 
(which is equal to the separation between the valence bands).
The band structure (\ref{bandstructure}) is calculated from the tight binding 
approach, where $v_F$ is connected to the nearest-neighbour hopping terms 
for electrons to move in each of the two graphene planes, and the distance 
between Carbon atoms in one monolayer (see Ref. \cite{Nicol2008}), whereas 
$\gamma$ is the hopping parameter corresponding to electrons hoping from one 
layer to the other and vice versa \cite{Nicol2008}. 
The two graphene layers are stacked one above the other according to the 
so-called Bernal-type stacking (e.g., see Ref. \cite{CastroNeto}). 
We emphasize that the perfect nesting of bands gives rise to the stronger 
plasmon like features of TE plasmons in bilayer than in monolayer graphene. 
The four bands are illustrated in Fig. \ref{bands} along 
with some of the electronic transitions which result in absorption. 
Absorption depends on $\gamma$ and the Fermi level $\mu$; the latter can be 
changed by applying external bias voltage. 

The imaginary part of the conductivity can be calculated from 
$\Re\sigma(\omega)$ by using the Kramers-Kronig relations 
\begin{equation}
\Im\sigma(\omega)=-\frac{2\omega}{\pi}
{\cal P}\int_{0}^{\infty}
\frac{\Re\sigma(\omega')}{\omega'^2-\omega^2}d\omega',
\end{equation}
which yields
\begin{eqnarray}
\frac{\Im\sigma(\omega)}{\sigma_0} 
& = & f(\Omega,2\mu)+g(\Omega,\mu,\gamma) \nonumber \\
& + & [f(\Omega,2\gamma)+g(\Omega,\gamma,-\gamma)]\Theta(\gamma-\mu) \nonumber \\
& + & [f(\Omega,2\mu)+g(\Omega,\mu,-\gamma)]\Theta(\mu-\gamma) \nonumber \\
& + & \frac{\gamma^2}{\Omega^2}
\left[ \frac{\Omega}{\pi(2\mu+\gamma)}+f(\Omega,2\mu+\gamma) \right] \nonumber \\
& + & \frac{\gamma^2}{\Omega^2}
\left[ \frac{\Omega}{\pi\gamma}+f(\Omega,\gamma) \right]\Theta(\gamma-\mu) \nonumber \\
& + & \frac{\gamma^2}{\Omega^2}
\left[ \frac{\Omega}{\pi(2\mu-\gamma)}+f(\Omega,2\mu-\gamma) \right]\Theta(\mu-\gamma) \nonumber \\
& + & \frac{a(\mu)}{\pi\Omega}+\frac{2 \Omega b(\mu)}{\pi(\Omega^2-\gamma^2)},
\label{ImS}
\end{eqnarray}
where
\begin{eqnarray}
f(x,y) & = & \frac{1}{2\pi} \log\left| \frac{x-y}{x+y} \right|,  \nonumber \\
g(x,y,z) & = & \frac{z}{2\pi} 
\frac{(x-z)\log|x-2y|+(x+z)\log|x+2y|-2x \log|2y+z|}
{x^2-z^2} , \nonumber \\
a(\mu) & = & \frac{4\mu(\mu+\gamma)}{2\mu+\gamma}+
\frac{4\mu(\mu-\gamma)}{2\mu-\gamma}\Theta(\mu-\gamma) , \nonumber \\
b(\mu) & = & \frac{\gamma}{2} \left[\log\frac{2\mu+\gamma}{\gamma} -
\log\frac{2\mu-\gamma}{\gamma}\Theta(\mu-\gamma) \right],
\label{ImS1}
\end{eqnarray}
$\sigma_0=e^2/2\hbar$, $\Theta(x)=1$ if $x\geq 0$ and zero otherwise, and $\Omega=\hbar\omega$. 
Here we assume zero temperature $T\approx 0$, which is a good approximation for sufficiently 
doped bilayer graphene where $\mu \gg k_B T$.
Formulae (\ref{ImS}) and (\ref{ImS1}) are used to describe the properties of 
TE plasmons.

\begin{figure}[htbp]
\centerline{
\mbox{\includegraphics[width=0.44\textwidth]{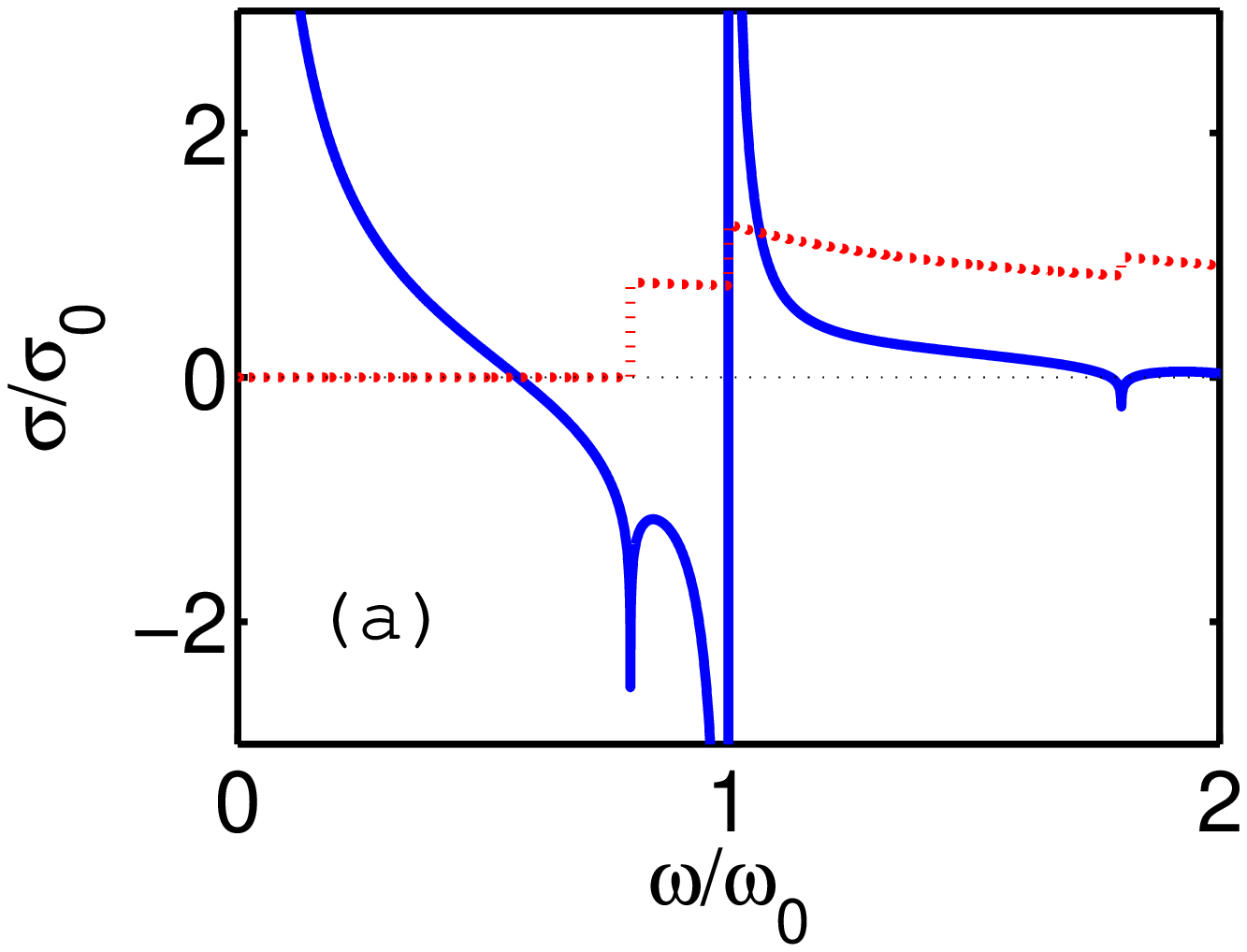}}
\mbox{\includegraphics[width=0.44\textwidth]{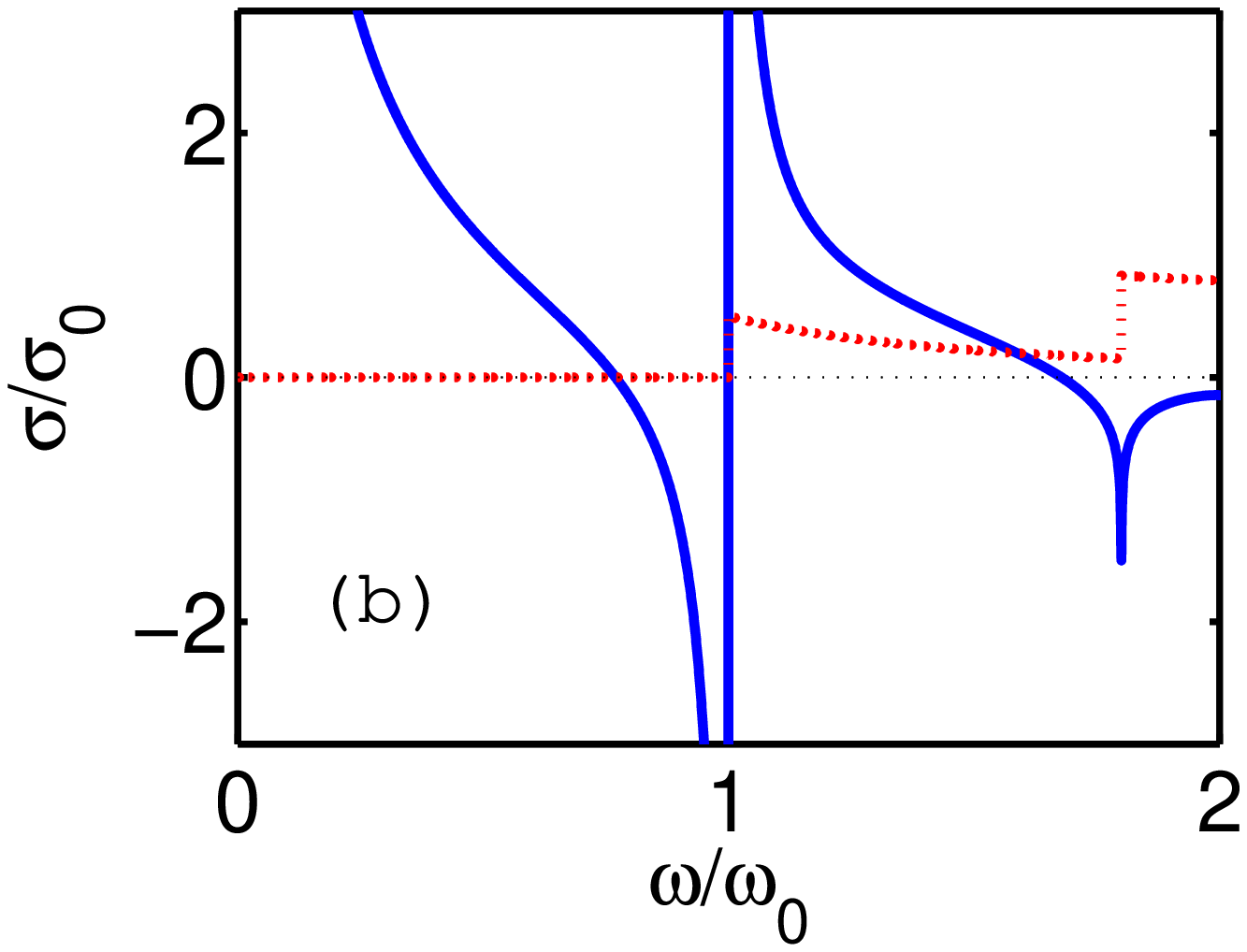}}
}
\caption{
The real (red dotted lines) and imaginary (blue solid lines) part of the 
conductivity of bilayer graphene for two values of doping: 
$\mu=0.4\gamma$ (a), and $\mu=0.9\gamma$ (b). 
The conductivity is in units of $\sigma_0=e^2/2\hbar$, and the 
frequency is in units of $\omega_{0}=\gamma/\hbar$.
The $\delta$-functions in $\Re\sigma(\omega)$ at $\omega=0$ (intraband transitions)
and $\omega=\gamma/\hbar$ (transitions from the lower to the upper conduction band
depicted as green solid arrows in Fig. \ref{bands}) are not shown
(see \cite{Nicol2008}).
}
\label{Sig}
\end{figure}

In Figure \ref{Sig} we show the real and imaginary part of the 
conductivity for two different values of the Fermi level:
$\mu=0.4\gamma$ and $\mu=0.9\gamma$ (we focus on the electron doped system $\mu>0$). 
Because plasmons are strongly damped by interband transitions, 
it is instructive at this point to discuss the kinematical requirements 
for the excitation of electron-hole pairs. 
If the doping is such that $\mu<\gamma/2$,
a quantum of energy $\hbar\omega$ (plasmon or photon) with in-plane momentum 
$q=0$ can excite an electron-hole pair only if $\hbar\omega>2\mu$ (excitations from the 
upper valence to the lower conduction band shown as red dot-dashed line in 
Fig. \ref{bands}). 
If $\mu>\gamma/2$, the $(q=0,\omega)$-quantum can excite an electron-hole pair only for 
$\hbar\omega\geq \gamma$ (excitations from the lower to the upper conduction band
shown as green solid lines in Fig. \ref{bands} occur at $\hbar\omega=\gamma$). 
If the plasmon/photon has in-plane momentum $q$ larger than zero, then 
interband transitions are possible for smaller frequencies
(see blue dashed lines in Fig. \ref{bands}). 
There is a region in the $(q,\omega)$-plane where electron-hole excitations 
are forbidden due to the Pauli principle (e.g., see figures in Refs. 
\cite{Borghi2009,Wang2010,Sensarma2010}). 
Because plasmons are strongly damped by these interband transitions 
(this is Landau damping), in our search for the 
TE plasmons, we focus on their dispersion curve in the regime where 
electron-hole pair formation is inadmissible (via first-order transition). 

\begin{figure}[htbp]
\centerline{
\mbox{\includegraphics[width=0.44\textwidth]{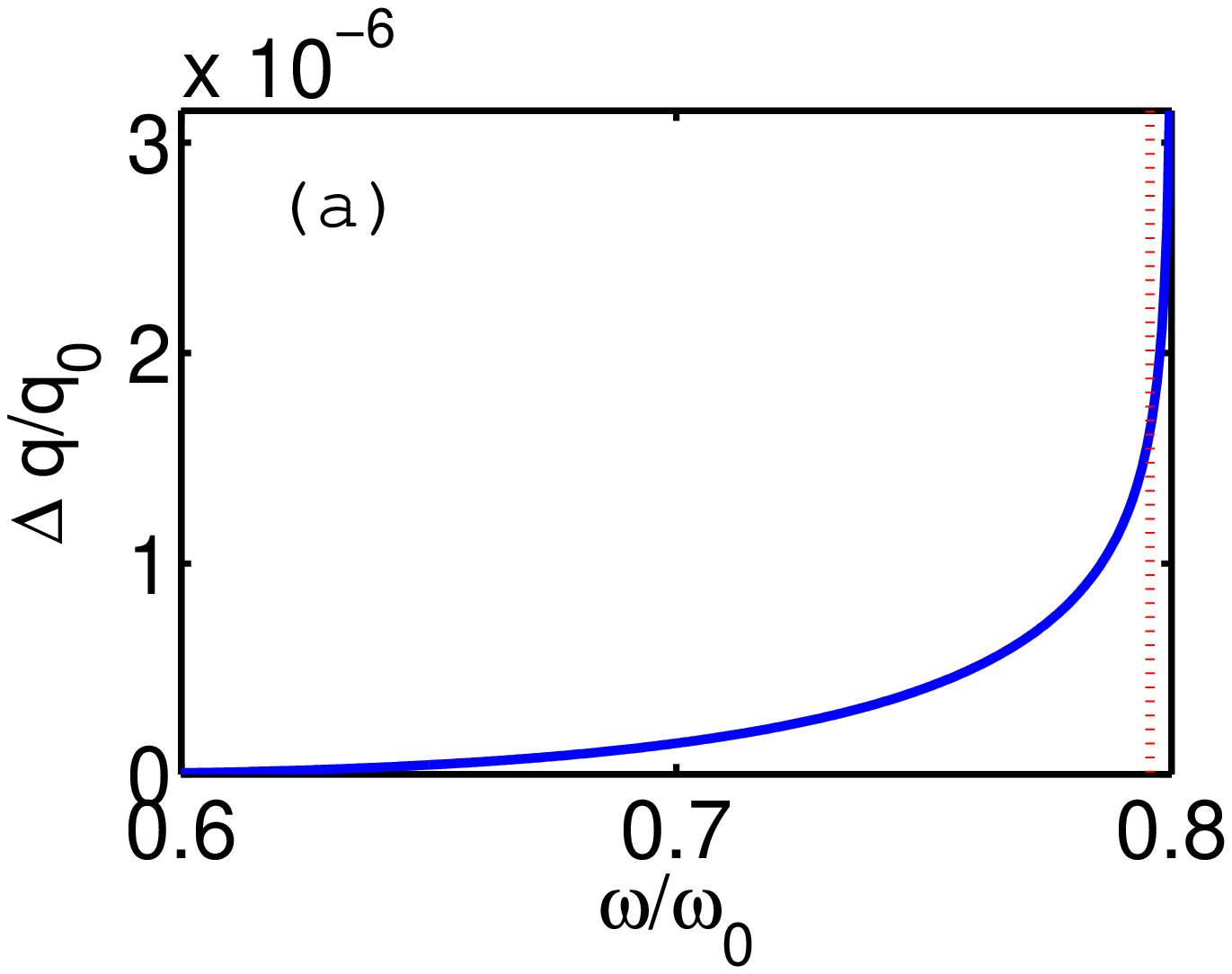}}
\mbox{\includegraphics[width=0.44\textwidth]{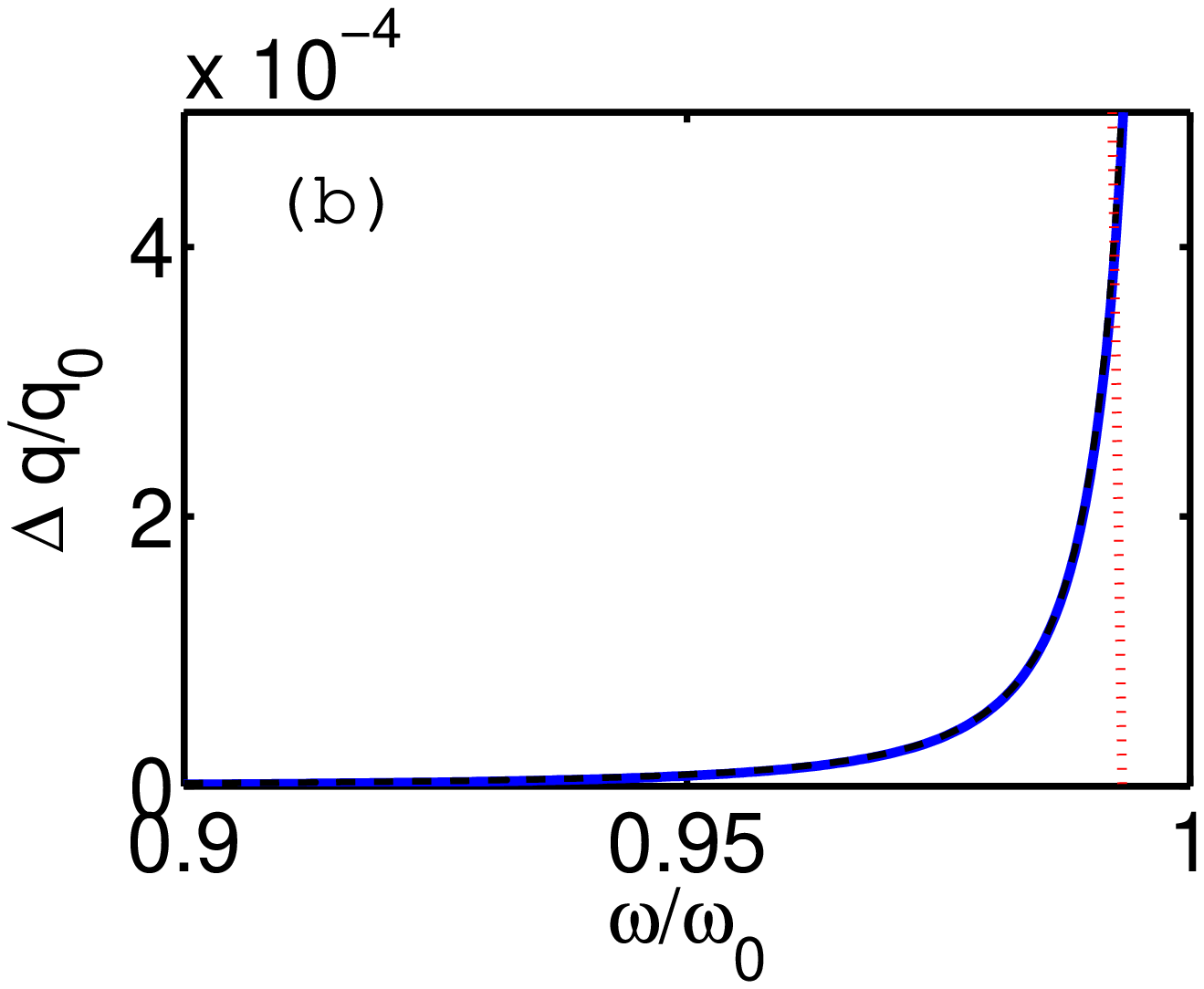}}
}
\caption{
The plasmon dispersion curve $\Delta q=q-\omega/c$ vs. $\omega$ 
for $\mu=0.4\gamma$ (a), and $\mu=0.9\gamma$ (b) is shown as blue solid 
line. To the right of the vertical red dotted lines plasmons can 
be damped via excitation of electron-hole pairs, whereas to the left of 
this line these excitations are forbidden due to the Pauli principle. 
Black dashed line in (b) (which closely follows the blue line) 
corresponds to Eq. (\ref{approxgamma}). 
The wave vector is in units of $q_0=\gamma/\hbar v_F$, and the 
frequency is in units of $\omega_{0}=\gamma/\hbar$. 
}
\label{Dq}
\end{figure}

In Figure \ref{Dq} we show the plasmon dispersion curves for 
$\mu=0.4\gamma$ and $\mu=0.9\gamma$; in the spirit of Ref. 
\cite{Mikhailov2007}, we show $\Delta q=q-\omega/c$ as a function 
of frequency $\omega$. 
Plasmons are very close to the light line and thus one can 
to a very good approximation write the dispersion curve as 
\begin{equation}
\Delta q\approx \frac{\omega}{8\epsilon_0^2c^3}\Im\sigma(\omega)^2.
\end{equation}
To the left (right) of the vertical red dotted line in Fig. \ref{Dq}, 
plasmon damping via excitation of electron-hole pairs is (is not) forbidden.  

For $\mu=0.4\gamma$, $\Im\sigma(\omega)$ is smaller than zero 
for $\omega$ in an interval of frequencies just below $2\mu$. 
From the leading term in $\Im \sigma (\omega)$ we find that 
departure of the dispersion curve from the light line is logarithmically slow: 
$\Delta q_{0<\mu<\gamma/2}\propto [\log|\hbar\omega-2\mu|]^2$. 
The same type of behavior occurs in monolayer graphene \cite{Mikhailov2007}.

However, for $\mu=0.9\gamma$, one can see the advantage of bilayer over
monolayer graphene in the context of TE plasmons. 
The conductivity $\Im\sigma(\omega)$ is smaller than zero in an interval 
of frequencies below $\gamma$. In this interval, 
the most dominant term to the conductivity is the last one from 
Eq. (\ref{ImS}), that is,
\begin{equation}
\Delta q_{\gamma/2<\mu<\gamma}\approx
\frac{\omega\sigma_0^2 }{2\pi^2\epsilon_0^2c^3}
\left[\frac{\hbar\omega b(\mu)}{\gamma^2-(\hbar\omega)^2}\right]^2.
\label{approxgamma}
\end{equation}
This approximation is illustrated with black dashed line in Fig. \ref{Dq},
and it almost perfectly matches the dispersion curve. 
Note that the singularity in $\Im\sigma(\omega)$ at $\hbar\omega=\gamma$ is  
of the form $1/(\gamma-\hbar\omega)$, whereas the singularity at $\hbar\omega=2\mu$
is logarithmic (as in monolayer graphene \cite{Mikhailov2007}). 
As a consequence, the departure of the dispersion curve from the 
light line in bilayer graphene is much faster for $\mu>\gamma/2$ than 
for $\mu<\gamma/2$, and it is faster than in monolayer graphene as well 
[note the two orders of magnitude difference between the abscissa scales 
in Figs. \ref{Dq}(a) and (b)]. 
Thus, we conclude that more pronounced plasmonic features of TE plasmons (shrinking of wave length 
which is measured as departure of $q$ from the light line) can be obtained 
in bilayer graphene. The term in $\Im\sigma(\omega)$ which is responsible 
for TE plasmons for $\mu>\gamma/2$ corresponds (via Kramers-Kronig relations) 
to the absorption term $b(\mu)\delta(\hbar\omega-\gamma)$ \cite{Nicol2008}, which arises from the 
transitions from the first to the second valence band
(shown as green solid arrows in Fig. \ref{bands}), which are perfectly nested 
and separated by $\gamma$. Thus, this unique feature of bilayer graphene 
gives rise to TE plasmons with more pronounced plasmon like features 
than in monolayer graphene. 

Before closing, let us discuss some properties and possible observation of TE plasmons. 
First, note that since the electric field oscillations are 
both perpendicular to the propagation vector ${\bf q}$, and 
lie in the bilayer graphene plane, the electric current ${\bf j}=\sigma(\omega) {\bf E}$
is also perpendicular to ${\bf q}$. Thus, ${\bf j}\cdot {\bf q}=0$, and the 
equation of continuity yields that the charge density is zero 
(i.e., one has self-sustained oscillations of the current).
In order to excite plasmons of frequency $\omega$ with light of the same 
frequency, one has to somehow account for the conservation of the momentum 
which is larger for plasmons. Since the momentum mismatch is relatively 
small, the standard plasmon excitation schemes such as the 
prism or grating coupling methods (e.g., see \cite{Barnes2003} and 
references therein) could be used for the excitation of these plasmons. 

In conclusion, we have predicted the existence of transverse electric (TE) 
plasmons in bilayer graphene. Since they exist very close to 
the light line, these plasmons are expected to show strong polariton 
character, i.e., mixing with photon modes. 
However, due to the perfectly nested valence bands of bilayer graphene, 
their dispersion departs much more from the light line than in monolayer 
graphene. 

We acknowledge most useful comments from Guy Bartal and Mordechai Segev 
from the Technion, Israel. 
This work was supported in part by the Croatian Ministry of Science 
(Grant No. 119-0000000-1015), the Croatian-Israeli scientific cooperation program
funded by the Ministries of Science of the State of Israel and the Republic of Croatia.
This work is also supported in part by the MRSEC program of National 
Science Foundation of the USA under Award No. DMR-0819762. M.S. was also 
supported in part by the S3TEC, an Energy Frontier Research Center funded 
by the U.S. Department of Energy, Office of Science, Office of Basic 
Energy Sciences under Award No. DE-SC0001299.


\end{document}